\documentclass[english,aps, manuscript]{revtex4}
\usepackage{graphicx}

\makeatletter

 \newenvironment{lyxlist}[1]
   {\begin{list}{}
     {\settowidth{\labelwidth}{#1}
      \setlength{\leftmargin}{\labelwidth}
      \addtolength{\leftmargin}{\labelsep}
      }}
   {\end{list}}

\usepackage{babel}
\makeatother
\begin{document}

\preprint{Preprint}

\title{Finite Quantum Grand Canonical Ensemble and Temperature from Single Electron Statistics in a Mesoscopic Device}

\author{Enrico Prati}

\affiliation{Laboratorio Nazionale Materiali e Dispositivi per la Microelettronica,
Consiglio Nazionale delle Ricerche - Istituto Nazionale per la Fisica della Materia, Via Olivetti 2, I-20041
Agrate Brianza, Italy}

\email{enrico.prati@cnr.it}

\begin{abstract}
I present a theoretical model of a quantum statistical ensemble for which, unlike in conventional physics, the total number of particles is extremely small. The thermodynamical quantities are calculated by taking a small $N$ by virtue of the orthodicity of canonical ensemble. The finite quantum grand partition function of a Fermi-Dirac system is calculated. The model is applied to a quantum dot coupled with a small two dimensional electron system.  Such system consists of an alternatively single and double occupied electron system confined in a quantum dot, which exhanges one electron with a small $N$ two dimensional electron reservoir. The analytic determination of the temperature of a $(1\leftrightarrow 2)$ electron system and the role of ergodicity are discussed.  The generalized temperature expression in the small $N$ regime recovers the usual temperature expression by taking the limit of $N\rightarrow\infty$ of the electron bath.
\end{abstract}

\newpage

\maketitle

\section{Introduction}
A deep connection exists between the nature of the elementary objects described by quantum mechanics, and the emerging properties of thermodynamical quantities. Bohr clarified that the complementarity principle should apply to energy and temperature measurements. Indeed, the determination of the former is incompatible with the knowledge about the second, when they refer to elementary objects \cite{Rosenfeld61}. 
After the creation of solid state quantum dots \cite{Beenakker91}, it became possible to explore confined fermionic systems constituted of few electrons \cite{Sanquer00}, down to a single localized electron \cite{Rogge06,Koppens06,Prati08i,Prati09}. Such systems are electrically probed by means of accurate charge sensing capable to determine current fluctuations corresponding to a variation of charge much below the charge unit \cite{Ralls86,Prati06,Prati08,Prati08ii}.
The experimental determination of the electron temperature in a nanostructure is a difficult task \cite{Giazotto06}, but it becomes a principle issue when the system consists of a $(1\leftrightarrow2)$ electron system in particle exchange with a few electron bath. Even if the definition of temperature $T=\left( \frac{\delta S}{\delta U} \right)^{-1}$ in terms of the energy $U$ and the entropy $S$ holds independently from the size of the ensemble \cite{Landau}, the systems here considered are far from the conventional assumptions to derive thermodynamics from statistical physics. Apparently, it should not be possible to associate a temperature to such small systems. 
The reduction of the size of the system down to one electron in thermal and particle exchange with a finite electron system (small number of electrons $N$) implies a twofold change of perspective to determine the thermodynamical quantities. The first major change is the shift from space to time ensembles. Statistics can be recovered on a few particle open system only by considering the average of measureable quantities in the time domain. The second change consists in the generalization of thermodynamics by removing the limit of large $N$, which implies that the terms at the order $1/N$ are relevant in the determination of the physical quantities. The equivalence under the first change of perpective is granted by the ergodicity, while the second change is granted by the orthodicity. 
In the Section II, I define the finite quantum grand partition ensemble in the limit of small $N$.  Next, I determine the occupation probability of an electronic system capable to contain either one or two electrons. The state equation is derived and it surprisingly depends on the heat capacity per area unit $c_A$. In Section III, the determination of the generalized temperature expression and its dependence from experimentally measurable quantities is described. Such evaluation is done by considering the realistic experimental condition of a single electron quantum dot coupled with a small electron reservoir. The results are summarized and briefly discussed in the Conclusions.

\section{Finite quantum grand canonical ensemble of a $(1\longleftrightarrow 2)$ fermionic system}

The grand canonical ensemble consists of a large open ensemble made of identical systems, which is in thermal equilibrium with a reservoir at a given temperature. The ensemble under investigation and the thermal reservoir may exchange energy and particles. In this section the statistical physics of a similar system is investigated. The model is derived for a generic system of fermions and it applies to electrons in a quantum dot. Contrarily to the grand canonical ensemble, here the open system under investigation may contain only 1 or 2 fermions. The reservoir is a low dimensional system of few fermions.
According to the physics and technology of semiconductor nanodevices, to which the present study can be applied, the reservoir is well approximated with a two dimensional electron system (2DES), with a negligible extension in the direction perpendicular to the plane of the electrons \cite{Simmons07,Prati08i}. To grant the uniformity of the notation, the island capable to capture one extra electron is exemplified with a trap spatially extended along two dimensions, without any loss of generality. Such assumption is realistic, being the confinement of the electron wavefunction symmetric in the $x-y$ plane both in defects at the $Si/SiO_2$ interface and in a quantum dot which is either litographically or split gate defined in a heterostructure \cite{Simmons07}.
Therefore, all the definitions in the following refer to surfaces $\Sigma_i$ instead of volumes $V_i$ where $i=1$ indicates the quantum dot and $i=2$ a thermal bath made of $N_2$ electrons. 

A system is orthodic if the averages of the physical quantities on a given distribution recover the laws of thermodynamics. The canonical ensemble is natively orthodic so such properties is granted for the finite $N$ grand canonical ensemble by construction. 

Said $H(N)$ the Hamiltonian of the system constitued by $N$ electrons distributed between the island and the 2DEG, it can be separated by $H=H_1(N_1)+H_2(N_2)$ where $N_{1} = 1,2$. 

Consequently one defines the finite quantum grand partition function for identical particles:

\begin{equation}
Z(\Sigma,N,T)=\sum_{N_1=1}^{2} Z(\Sigma_1,N_1,T) Z(\Sigma_2,N_2,T)=
\end{equation}

\begin{equation}
=\sum_{N_1=1}^{2} tr \left(e^{-\beta H_1}\right) tr \left( e^{-\beta H_2}\right)
\end{equation}

where $N=N_1+N_2$ and $\Sigma=\Sigma_1+\Sigma_2$. It is introduced the function 
\begin{equation}
\rho(N_1)=\frac{Z(\Sigma_2,N_2,T)}{Z(\Sigma_1+\Sigma_2,N_1+N_2,T)}e^{-\beta H_1(N_1)}
\end{equation}

where $\beta=\left( kT \right)^{-1}$ which by definition satisfies:
\begin{equation}
\sum_{N_1=1}^{2} tr\left(  \rho(N_1) \right) =1
\end{equation}

Since $tr\left(\rho\right)$ is the canonical partition function times $Z(\Sigma_2,N_2,T)/Z(\Sigma,N,T)$, in the following such ratio is calculated by means of the Helmholtz potential $\Psi=- \beta^{-1} log Z$, which gives:
\begin{equation}
\frac{Z(\Sigma_2,N_2,T)}{Z(\Sigma,N,T)}=e^{\beta\Psi(\Sigma_1+\Sigma_2,N_1+N_2,T)-\beta\Psi(\Sigma_2,N_2,T)}
\end{equation}

In order to explicitly evaluate $\Psi$, the internal energy $U$ is now calculated, since $\Psi=U-TS=U+T \frac{\delta \Psi}{\delta T}$ where $S=-\left( \frac{\delta \Psi}{\delta T} \right)$. Temperature $T$ is defined as usual from $T=\left(\frac{\delta S}{\delta U}\right)^{-1}$.
For a Fermi 2DES of $M$ electrons in a surface with area $A$, in the limit of small temperature $T$, it holds:
\begin{equation}
U(M,A)=A \cdot u = A \left\{ \int_0^{\mu_F} Eg(E)dE+ \frac{\pi^2}{6}\left(k_B T \right)^2 \left[ \mu g'(\mu)+g(\mu)\right] \right\}
\end{equation}

\begin{equation}
M=A \cdot n = A \left\{ \int_0^{\mu_F} g(E)dE+ \frac{\pi^2}{6}\left(k_B T \right)^2 \left[ g'(\mu)\right] \right\}
\end{equation}

where the density of states per surface unit is $g(E)dE=\frac{m }{\pi^2 \hbar^2}dE$ for a two dimensional system with $d=2$ and $\mu_F$ is the Fermi energy. Since $g'(E)=0$ at $d=2$,
\begin{equation}
U(M, A)=\frac{g}{2} A \mu^{2}_{F}+ \frac{\pi^2}{6} gA (k_B T)^2
\end{equation}
\begin{equation}
M=g A \mu_F 
\end{equation}
The inversion of Equation 9 gives the chemical potential
\begin{equation}
\mu(M,A)=\frac{M}{g A}
\end{equation}
so the internal energy $U$ is equivalently written
\begin{equation}
U(M)=\frac{1}{2} \frac{M^2}{gA}+ \frac{\pi^2}{6} gA (k_B T)^2
\end{equation}

The above relations are valid for $M=N_2$ electrons with $A=\Sigma_2$ and for $M=N_1+N_2$ with $A=\Sigma_1+\Sigma_2$, while $U(N_1)$ is treated separately because of the dependence of the nature of confinement when $N_1=1,2$.
The polynomial shape of $U(M)=a_{M,A}+b_{A}T^2$, where $a_{M,A}=\frac{1}{2} \frac{M^2}{gA}$ and $b_A = \frac{\pi^2}{6} gA (k_B)^2$ as a function of the temperature $T$ implies that $\Psi(M,A)=a_{M,A}-b_{A}T^2$ so

\begin{equation}
\Psi(M,A)=\frac{1}{2} \frac{M^2}{gA}- \frac{\pi^2}{6} gA (k_B T)^2
\end{equation}

It is useful to calculate the heat capacity per area unit at constant  surface
\begin{equation}
c_{A}=\frac{1}{A}\left(\frac{\partial U(M,A)}{\partial T}\right)_A= \frac{\pi^2}{3} g k_B^2 T 
\end{equation}

and the pressure
\begin{equation}
P(M,A)=-\left(\frac{\partial \Psi(M,A)}{\partial A}\right)_T= \frac{1}{2} \frac{M^2}{g A^2} + \frac{\pi^2}{6} g (k_B T)^2 
\end{equation}

Since the island which traps the electron has a very small spatial extension, it holds the approximation $\Sigma \cong \Sigma_2$ which simplifies the analysis.
It is now possible to evaluate the Helmholtz energy variation between $N_1+N_2$ and $N_2$ electrons:

\begin{equation}
\Delta \Psi= \Psi(N_1+N_2,\Sigma_1+\Sigma_2)-\Psi(N_2,\Sigma_2)= 
\end{equation}
\begin{equation}
=\frac{(N_1 + N_2)^2}{2 g (\Sigma_1 + \Sigma_2)} - \frac{N_2^2}{2 g \Sigma_2} -
\frac{\pi^2}{6} g (\Sigma_1 + \Sigma_2) (k_B T)^2 + \frac{\pi^2}{6} g \Sigma_2 (k_B T)^2 \cong
\end{equation}
\begin{equation}
\cong  \frac{N_1^2 + 2 N_1 N_2 + N_2^2}{2 g \Sigma_2} - \frac{N_2^2}{2 g \Sigma_2} -
\frac{\pi^2}{6} g \Sigma_1  (k_B T)^2 =
\end{equation}
\begin{equation}
=N_1 \mu(N_2,\Sigma_2) \left( 1 + \frac{N_1}{2N_2} \right) - \frac{1}{2} c_A \Sigma_1 T
\end{equation}

$\Delta \Psi$ goes in the Eq. 5 and gives
\begin{equation}
\frac{Z(\Sigma_2,N-N_1,T)}{Z(\Sigma,N,T)}=e^{\beta\mu(N_2) N_1\left(1+\frac{N_1}{2N_2}\right) - \frac{c_A \Sigma_1}{2 k_B}}
\end{equation}

The \textit{finite quantum grand canonical ensemble} can be therefore defined by the pair $(\mathcal{E},\rho)$ with $\mathcal{E}=\Gamma$ and:

\begin{equation}
\rho(N_1)= z^{N_1 (1+N_1/2N_2)}e^{- \frac{c_A \Sigma_1}{2 k_B} -\beta H(N_1)} 
\end{equation}

where $z=e^{\beta \mu(N_2)}$, while $\Gamma$ indicates all the states of the system over which the summations are performed and it consists of the discrete quantum analogue of the Gibbs $\Gamma$-space.\cite{Huang63}
 
The probability of occupation of the subsystem 1 with $L$ fermions is therefore

\begin{equation}
p(L)=\frac{tr \rho(L)}{\sum_{N_1=1}^2 tr \left(\rho(N_1)\right)}=\frac {z^{L(1+L/2N_2)} tr \left(e^{-\beta H_1(L)}\right)}
 {\sum_{N_1=1}^2 z^{N_1 (1+ N_1 /2 N_2)} tr \left(e^{-\beta H_1(N_1)}\right)}=
\end{equation}
\begin{equation}
=\frac {z^{L(1+L/2N_2)} tr \left(e^{-\beta H_1(L)}\right)}
 {Z_{QG}}
\end{equation}

where
\begin{equation}
Z_{QG}=\sum_{N_1=1}^2 z^{N_1 (1+N_1/2N_2)}  tr \left( e^{-\beta H_1(N_1)}\right)
\end{equation}

The present section is concluded by expressing the state equation and the relationship between the temperature $T$ and the occupation statistics. Since $\sum_{N_1=1}^2 tr \left(\rho(N_1)\right)=1$, it holds: 
\begin{equation}
Z_{QG} \cdot e^{-\frac{c_A \Sigma_1} {2 k_B}}=1
\end{equation}

or equivalently, if one considers the logarithm of both sides: 

\begin{equation}
log Z_{QG} = +\frac{c_A \Sigma_1} {2 k_B}
\end{equation}

The generalized temperature of a $\left(1\leftrightarrow 2\right)$ system is therefore given by inverting

\begin{equation}
\frac{p(1)}{p(2)}=  \frac{z^{1+\frac {1}{2N_2}} tr (e^{-\beta H_1(1)})} {z^{2 \left(1+\frac {1}{N_2}\right)} tr (e^{-\beta H_1(2)})}
\end{equation}

In the next section such ratio is explicitly evaluated by considering the realistic value of a possible solid state quantum device.

\section{Single electron temperature in a quantum dot with a small $N$ electrons reservoir}

In this section the physical parameters involved in the electron occupation probability of a realistic quantum dot and the consequent experimental determination of the generalized temperature are discussed.
The study of the electron occupation of a quantum dot can be realized by measuring its charge state by the current in a channel electrically coupled to the electron charges confined in the dot \cite{Prati08ii,Fricke07}. In the case of a natural quantum dot like a donor or a lattice point defect close to the $Si/SiO_2$ interface, the channel is provided by the two dimensional gas formed at the interface by applying a gate voltage \cite{Ralls86,Prati08}. In the case of lithographically defined quantum dots, a current flows in the proximity of the island which confines the localized charges \cite{Simmons07}.
Let's consider the simplest system as possible, like a point defect close to the $Si/SiO_2$ interface. Typically the 2DES is confined in a $(50-300)\times (50-300) nm^2$, while the electron wavefunction is spread along few nm in the direction perpendicular to the 2DES \cite{Palma97}.
The point defect can only accept one extra electron. When the defect is paramagnetic (it switches from $N_1=1$ to $N_1=2$ and $\it viceversa$), the first electron fills the ground state of a hydrogen-like shell, at energy $E(1)=E_T$. Indeed, the high extraction energy of the unpaired electron makes impossible to achieve the ionization of the first localized electron, unless a metal gate is used to manually modify the charge state from $N_1=1$ to $N_1=0$, differently from the case here considered. 
Because of the spin degeneracy, a second electron can be captured at the same energy level $E_T$ to constitute a singlet state $\frac{1}{\sqrt{2}}\left(\left|\uparrow \right\rangle  \left| \downarrow \right\rangle - \left| \downarrow \right\rangle \left|\uparrow \right\rangle\right) $, with two extra contributions due to the Coulomb charging energy $\Delta E_C$ and to the lattice relaxation $\Delta E_L$. While the origin of the first contribution is straightforward, the second requires a short discussion. The presence of the second electron involves indeed a rearragement of the lattice \cite{Henry77}. At the low temperature here considered the 2DES is weakly coupled with the crystal \cite{Kivinen03}. Phonons are involved in the emission and capture of one electron from the defect and the relaxation of the crystal implies a change of energy of $S_{HR} \hbar \omega$ where $S_{HR}$ is the Huang-Rhys factor and $\omega$ is the average phonon frequency in the configuration coordinate picture \cite{Goguenheim90}. In the following we consider $\Delta E_L = S_{HR} \hbar \omega$ the energy gain of the lattice when an electron is captured. 
In a natural quantum dots it is not possible to capture a third electron because the energy of the system would exceed the conduction band edge energy.

The total energy of the two electrons is therefore

\begin{equation}
E(2)=2E_T + \Delta E_C + \Delta E_L
\end{equation}

Experimentally, the time resolved trace of the current is analyzed by means of fast digitizers and they typically appear as a random telegraph signal for both the quantum dots \cite{Fricke07} and the point defects \cite{Prati06}. A typical switching current trace is shown in Figure 1. The capture time $\tau(1)$ and the emission time $\tau(2)$ are obtained from the average of thousend of switching events and they generally obey to a Poissonian statistics. Deviations from such statistics are quantified by a non unit Fano factor and they reflect quantum coherence typical of pecularly coupled quantum dots \cite{Kiesslich07}.
Their ratio is governed by the ratio between the occupation probability between the two occupation probabilities:

\begin{equation}
\frac{\tau(1)}{\tau(2)}=\frac{p(1)}{p(2)}
\end{equation}

It is consequently suitable to calculate the relative occupation probability for the two states with 1 or 2 electrons, since their ratio can be experimentally accessed. For the particular case here discussed, when the subsystem 1 is a paramagnetic point defect

\begin{equation}
\frac{p(1)}{p(2)}= 2 e^{-\beta \mu \left( 1+ \frac{3}{2N_2} \right)}  \frac {e^{-\beta E_T}} {e^{-\beta 2 E_T + \Delta E_C + \Delta E_L}}=
\end{equation}
\begin{equation}
= 2 e^{\beta \left( E_T+\Delta E_L+\Delta E_C- \mu (N_2) \left( 1+ \frac{3}{2N_2} \right) \right)}  
\end{equation}

The factor 2 takes into account the spin degeneracy of the ground state.
The values of $p(1)$ and $p(2)$ are experimentally obtained as the average occupation time in the states 1 and 2 respectively. Such identification is possible by virtue of the ergodic hypothesis. The complete experimental determination of the parameters involved in the Eq. 30 provides the generalized temperature $T$ shared by the electrons in the quantum dot and the electrons in the 2DES. The generalized temperature of the electron(s) localized in the dots in thermal equilibrium with the small $N_2$ electron bath may consequently be defined:
\begin{equation}
T=k_B^{-1} ln \frac{2 \tau(2)}{\tau(1)} \left( E_T+\Delta E_L+\Delta E_C- \mu (N_2) \left( 1+ \frac{3}{2N_2} \right) \right)  
\end{equation}
It is remarkable that such definition of temperature of a time ensemble of few electrons concides with the usual one for a space ensable just by taking $N_2 \rightarrow\infty$. For this reason it can be considered a meaningful extension of the temperature definition for a small quantum system of electrons of the kind treated in the present paper. Such result can be easily adapted to the case of the confinement induced in a artificial quantum dot, which does not require the lattice relaxation term. 

\section{Conclusion}
The finite quantum grand canonical ensemble $(\mathcal{E},\rho)$ has been defined for a fermionic systems constituted by a cell capable to confine either 1 or 2 electrons, and a thermal bath of finite and small size made of a two dimensional system of few electrons. The state equation is governed by the heat capacity per surface unit $c_A$ instead of the pressure. Averaging on a time ensemble substitutes the average on space ensembles for the determination of the thermodynamical quantities. The ensemble has been applied to a quantum dot constituted by a natural point defect at the $Si/SiO_2$ interface and it holds for a general $(1\leftrightarrow2)$ system. The ratio of the characteristic times monitored by the current two-state fluctuations is given by the ratio between the occupation probability calculated with the presented approach. Therefore, the generalized temperature of such a small time ensemble can be defined and extracted from the ratio between the average characteristic times of the two current states. Such definition of temperature returns the usual temperature by taking the limit of $N\rightarrow\infty$ of the electron bath.

\begin{acknowledgments}
The author would like to thank Sergio Servadio (Universita di Pisa) for the careful reading of the manuscript and the useful suggestions and criticisms.
\end{acknowledgments}

\newpage

\section{Figure Captions}

\begin{lyxlist}{00.00.0000}
\item [Fig.1] The change of the charge occupation of both an artificial quantum dot and a natural point defect is monitored by recording the sudden changes of an ultra weak channel current electrically coupled with it. The capture and the emission phenomena are generally governed by a Poisson process which determines the average occupation time. $N_1$ is the number of electrons in the dot. In this example the high current state is associated to the $N_1 = 1$ occupation. The two states could also be reversed ($N_1=2$ would refer to the high current state) depending on the microscopic nature of the electrostatic coupling of the island with the two dimensional system.

\end{lyxlist}

\end{document}